\documentclass[manuscript]{aastex61}
\usepackage{graphicx}
\usepackage{epsfig}
\usepackage{color}

\begin{document}
\title{Hybrid stars in the light of GW170817}
\correspondingauthor{Prasanta Char}
\email{pchar@iucaa.in}

\author{Rana Nandi}
\affil{Tata Institute of Fundamental Research, Mumbai-400005, India}

\author{Prasanta Char}
\affiliation{Inter-University Centre for Astronomy and Astrophysics, Post Bag 4, Ganeshkhind, Pune - 411 007, India}

\begin{abstract}
We have studied the effect of tidal deformability constraint given by the binary neutron star merger event GW170817 on the 
equations of state (EOS) of hybrid stars. The EOS are constructed by matching the hadronic EOS described by relativistic
mean field (RMF) model and parameter sets NL3, TM1 and NL3$\omega\rho$ with the quark matter EOS described by modified MIT
bag model, via Gibbs' construction. It is found that the tidal deformability constraint along with the lower bound on maximum 
mass ($M_{\rm max}=2.01\pm0.04M_\odot$) significantly limits the bag model parameter space ($B_{\rm eff}^{1/4}$, $a_4$). 
We also obtain upper limits on the radius of $1.4M_\odot$ and $1.6M_\odot$ stars as $R_{1.4}\leq13.2-13.5$ km
and $R_{1.6}\leq13.2-13.4$ km, respectively for different hadronic EOS considered here.
\end{abstract}

\section{Introduction}
We have just entered the era of multi-messenger astronomy with the simultaneous detections of gravitational wave (GW) by the LIGO-VIRGO collaboration and its electromagnetic counterparts 
by $\sim 70$ ground and space based detectors \citep{Abbott2017}. All these observations strongly suggest that this event (GW170817) is associated with a system of binary neutron star (BNS) merger. 

While implications of this new data is being analyzed thoroughly in the context of the internal structure of a Neutron star (NS) \citep{Shibata2017,Bauswein2017,Rezzolla2018,Ruiz2018,Annala2017,Margalit2017,Zhou2017,Fattoyev2017,Radice2018,Paschalidis2017}, we are still far from definitive answers. It has been shown that with more events of this type, we will be able to constrain the NS parameters with better accuracy which in turn, will allow us to probe dense matter physics in precise detail \citep{Bose2018}. The equation of state (EOS) of matter has distinct effects on every stage of BNS evolution.
During the inspiral phase, the EOS-dependent tidal deformability parameter relates the quadrupolar response of a star with the external static tidal field exerted by its companion in the binary. Flanagan and Hinderer have shown how this parameter can influence the inspiral GW signal \citep{Flanagan2008,Hinderer2008,Hinderer2010}.
From the data of GW170817, LIGO-VIRGO collaboration \citep{Abbott2017} obtained an upper limit for the dimensionless tidal deformability of a $1.4M_\odot$ NS as $\Lambda(1.4M_\odot)\leq800$. The impact of this constraint has recently been studied on pure hadronic NS \citep{Fattoyev2017}, quark stars \citep{Zhou2017} as well as hybrid stars \citep{Paschalidis2017} having hadron-quark phase transition (PT) in the stellar interior. 

In this article we investigate the consequence of the tidal deformability constraints on the properties of hybrid stars. Unlike \citet{Paschalidis2017}, we construct EOS with PT from hadronic matter to quark matter via Gibbs' construction where the PT proceeds through the appearance of a quark-hadron mixed phase \citep{Glendenning1992, Glendenning2000}. The formation of mixed phase is favored when the surface tension between nuclear and quark matter is small ($\sigma\lesssim40$ MeV/fm$^2$) \citep{Alford2001,Voskresensky2003}. So far, the calculation of $\sigma$ is very much model dependent and can have values in the range $\sim 5-300$ MeV/fm$^2$ \citep{Alford2001, Voskresensky2003, Palhares2010,Pinto2012,Mintz2013,Lugones2013,Yasutake2014}. For higher values of $\sigma$ the phase transition is sharp and is treated with the prescription of Maxwell construction. As the value of $\sigma$ is not settled yet both the scenarios (Maxwell and Gibbs') are plausible. We adopt here the Gibbs' construction. In this case the outer part of the compact star contains hadronic matter, whereas the core can have either pure quark matter or the hadron-quark mixed phase - giving rise to a hybrid star.

The article is organized as follows. In the next section we present the details of EOS and the calculation of tidal deformability of hybrid stars. The effect of the tidal deformability constraint coming from GW170817 on the EOS of hybrid stars are investigated in Section \ref{sec:results}. Finally, we summarize and conclude in Section \ref{sec:summary}. 

We assume $G=c=1$ throughout this article, where $G$ and $c$ denote the gravitational constant and the speed of light, respectively.

\section{Set up}\label{sec:setup}
To construct the EOS of nuclear matter we adopt the relativistic mean-field (RMF) approach \citep{Walecka1974, Serot1979} where interaction between nucleons are described by the exchange of $\sigma$, $\omega$ and $\rho$ mesons. We use three widely used parameter sets: NL3 \citep{Lalazissis1997} which  includes non-linear self-interaction of $\sigma$-mesons, TM1 \citep{Sugahara1994} that has self-interacting terms for both $\sigma$ and $\omega$ mesons and NL3$\omega\rho$ \citep{Horowitz2001} where self-interaction of $\sigma$ mesons and coupling between $\omega$ and $\rho$ mesons are considered. It was shown in by \citet{Fortin2016} that the calculation of NS properties (especially radius) is unambiguous if the EOS is unified in the sense that the EOS of  crust and core are obtained using the same many-body theory. As $\Lambda$ is highly sensitive to radius ($\Lambda \sim R^5$), we take EOS of inner crusts from \citet{Grill2014} where they were calculated using the same parameters sets . For the outer crust we take DH EOS \citep{Haensel2007}. As most part of the outer crust is determined from the experimentally measured nuclear masses, the choice of outer crust does not have any significant impact on the observables.

For the description of quark matter we use the modified MIT bag model  described by the Grand potential \citep{Weissenborn2011}:
\begin{equation}
\Omega_{\rm QM} = \sum_i \Omega_i^0 + \frac{3\mu^4}{4\pi^2}(1-a_4)+B_{\rm eff},
\end{equation}
where $\Omega_i^0$ stand for the Grand potentials of non-interacting Fermi gases of up ($u$), down ($d$) and strange ($s$) quarks as well as electrons. The last two terms include the strong interaction correction and the nonperturbative QCD effects via two effective parameters $a_4$ and $B_{\rm eff}$, respectively with $\mu(=\mu_u+\mu_d+\mu_s$) being the baryon chemical potential of quarks.

We consider the PT from hadronic matter to quark matter via Gibbs' construction \citep{Glendenning1992}, where a mixed phase of hadronic and quark matter is formed between pure hadronic and quark phases. 

At the initial stage of the inspiral signal from a coalescing binary neutron star system, the tidal effects on a star can be taken at the linear order as,
\begin{equation}
Q_{ij}= - \lambda \mathcal{E}_{ij},
\end{equation}
where $Q_{ij}$ is the induced quadrupole moment of the star and $\mathcal{E}_{ij}$ can be assumed as external static tidal field exerted by the partner. The parameter $\lambda$ is related to the dimensionless, $\ell = 2$, electric-type tidal love number as,
\begin{equation}
\lambda = \frac{2}{3} k_2 R^5,
\end{equation}
where, $R$ is the radius of the star. We calculate $k_2$ and subsequently $\lambda$ following the framework developed by \citet{Hinderer2008,Hinderer2010}. 
However, in our present calculations, we use dimensionless tidal deformability $\Lambda$ (defined as $: \Lambda = \lambda/M^5$, where $M$ is the mass of the star) which is a direct observable quantity from the GW signal. The accumulated phase contribution due to the deformation from both the stars is imbibed in inspiral signal as the combined dimensionless tidal deformability which is given by ,
\begin{equation}
\tilde{\Lambda} = \frac{16}{13}\frac{\left(M_1 + 12 M_2\right)M_1^4 \Lambda_1 + \left(M_2 + 12 M_1\right)M_2^4 \Lambda_2}{\left(M_1 + M_2\right)^5},
\end{equation}
where, $\Lambda_1$ and $\Lambda_2$ are the individual defomability associated with the stars with masses $M_1$ and $M_2$ respectively \citep{Favata2014}.

\section{Results and discussion}\label{sec:results}
We construct the EOS of hybrid stars via Gibbs' construction as described in the previous section. For the hadronic part we take three different parameter sets NL3, TM1 and NL3$\omega\rho$ all of which give maximum NS masses ($M_{\rm max}$) more than $2M_\odot$ and are therefore compatible with the constraint of $M_{\rm max}=2.01\pm0.04M_\odot$ obtained from observation \citep{Antoniadis2013}. On the other hand, we generate a large number of quark matter EOS represented by the different values of $B_{\rm eff}^{1/4}$ and $a_4$. We then combine  all three hadronic EOS with all quark EOS via Gibbs' construction. However, we keep only those EOS where the starting density ($n_{\rm crit}$) of the mixed phase is larger than the crust-core
transition density ($n_{\rm cctd}$).

\begin{figure}
\centering
\begin{tabular}{cc}
\includegraphics[width=0.45\textwidth]{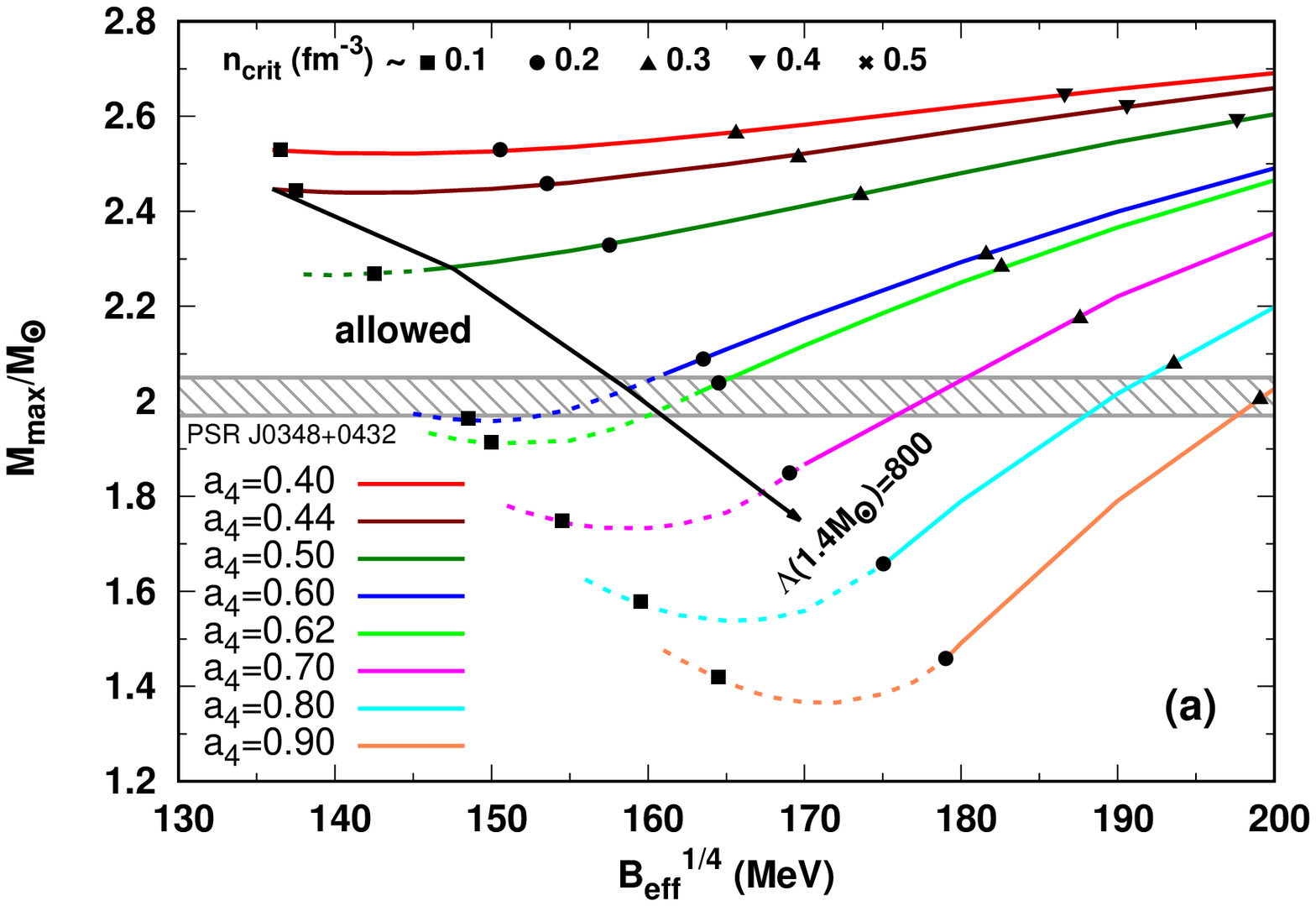}&
\includegraphics[width=0.45\textwidth]{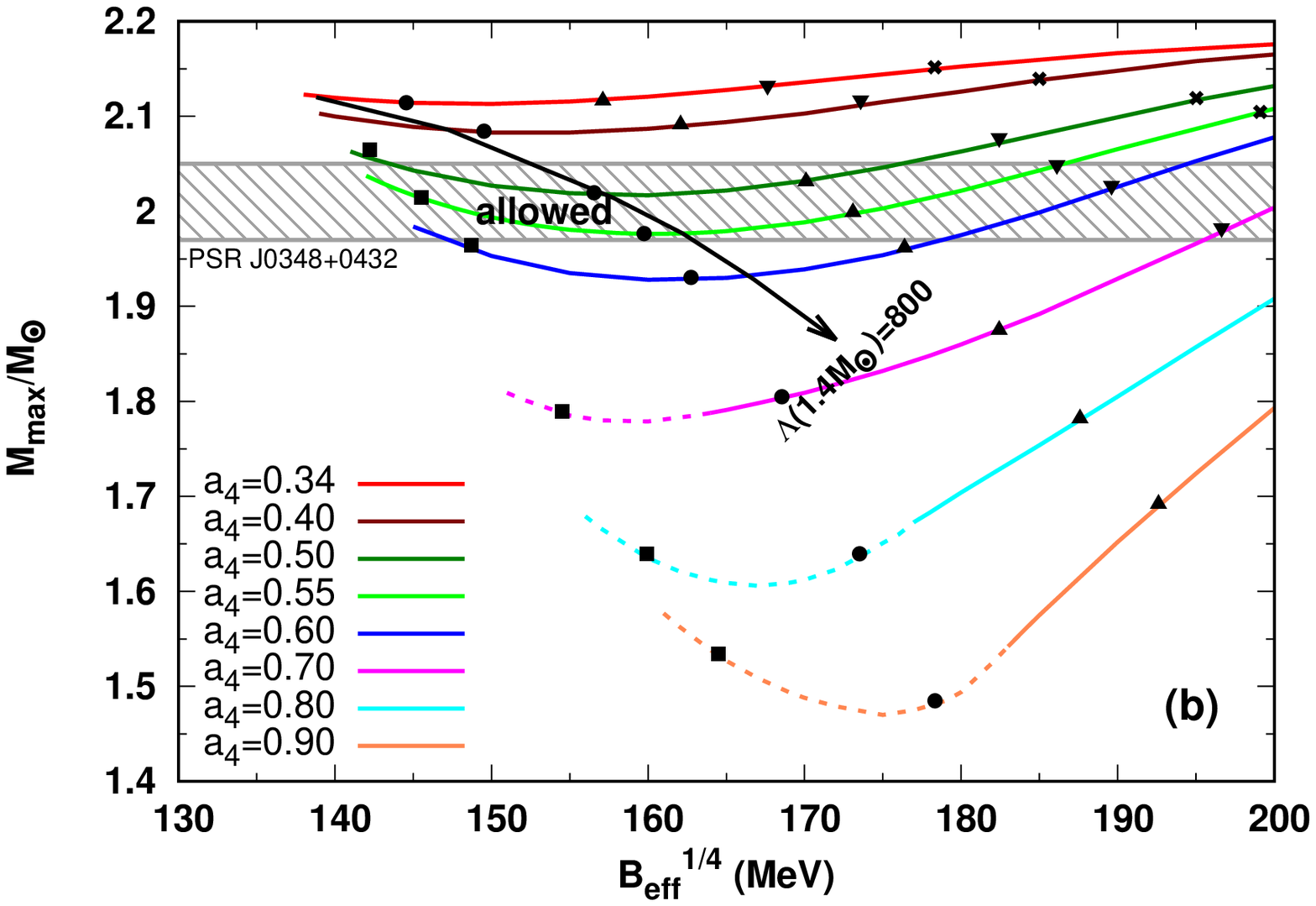}\\
\end{tabular}
\includegraphics[width=0.45\textwidth]{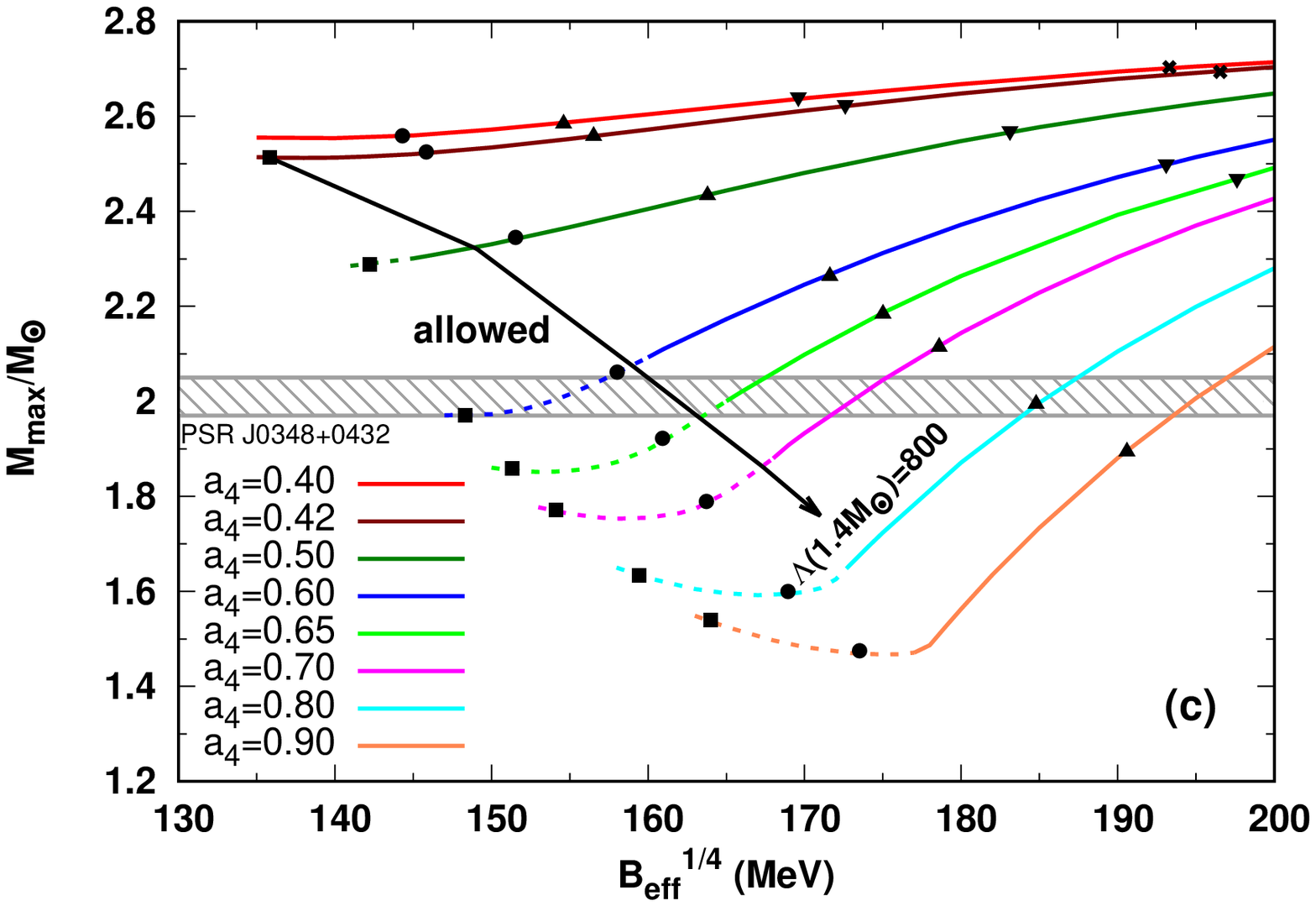}
\caption{Maximum masses and critical densities of hybrid stars as a function of $B_{\rm eff}^{1/4}$ and $a_4$ for (a) NL3, (b) TM1 and (c) NL3$\omega\rho$ parameter sets. Solid lines and dashed lines represent mixed phase core and pure quark matter core, respectively.}
\label{fig:Mmax}
\end{figure}
In Fig. \ref{fig:Mmax}, we show maximum masses and critical densities of hybrid stars for all EOS considered here. Similar figures were also obtained by \citet{Weissenborn2011} for NL3 and TM1 parameter sets. However, our figures for these two parameter sets are not identical to that of \citet{Weissenborn2011} at small $B_{\rm eff}^{1/4}$. This is because of the inclusion of the crust and excluding the EOS with $n_{\rm crit}\leq n_{\rm cctd}$. We also show the results for NL3$\omega\rho$ which were not presented earlier. For all three parameter sets we see that the observation of $\sim 2M_\odot$ NS already constraint the $(B_{\rm eff}^{1/4},a_4)$ parameter space, considerably.

\begin{figure}  
  \centering
  \begin{tabular}{cc}
    \includegraphics[width=0.45\textwidth]{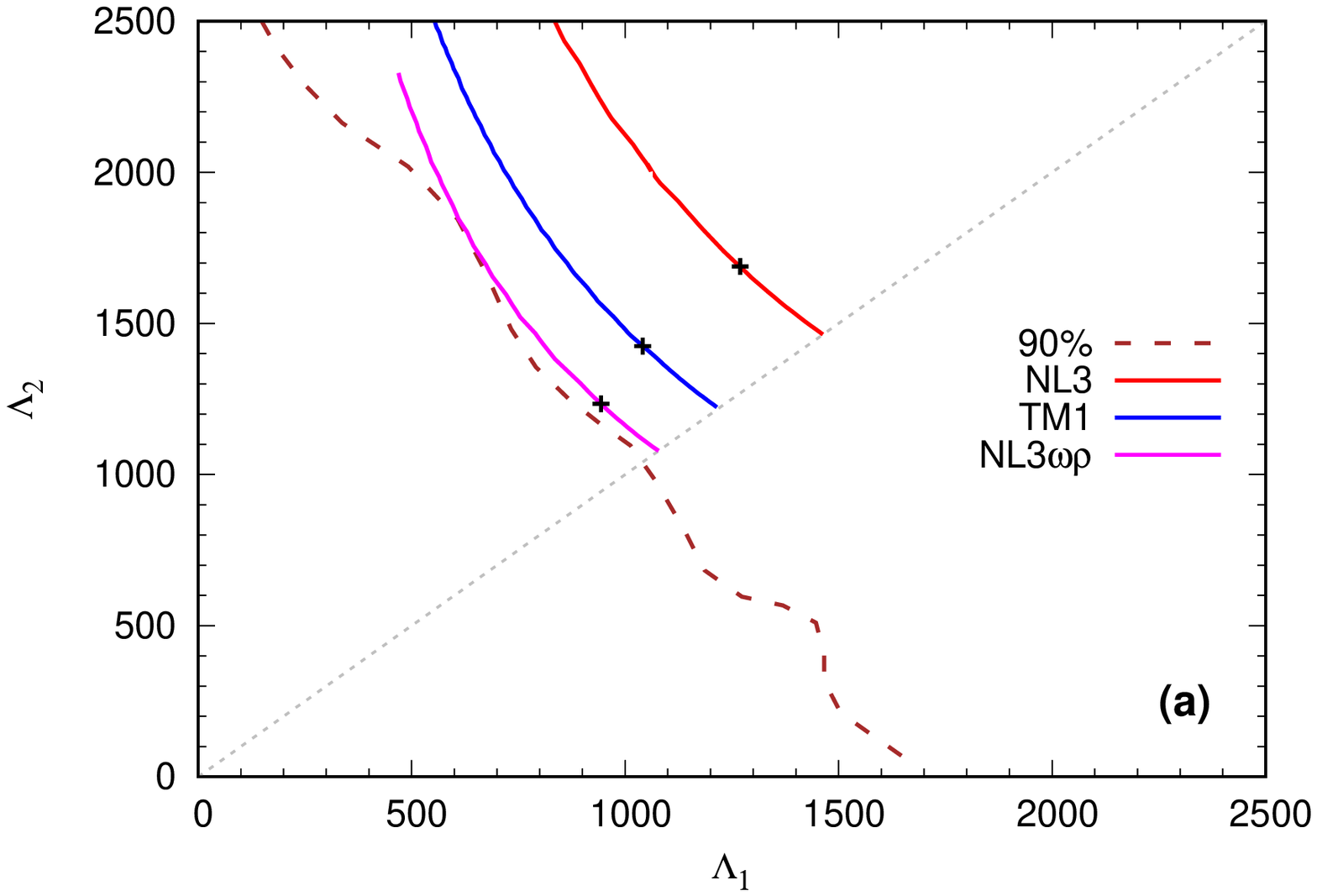}&
    \includegraphics[width=0.45\textwidth]{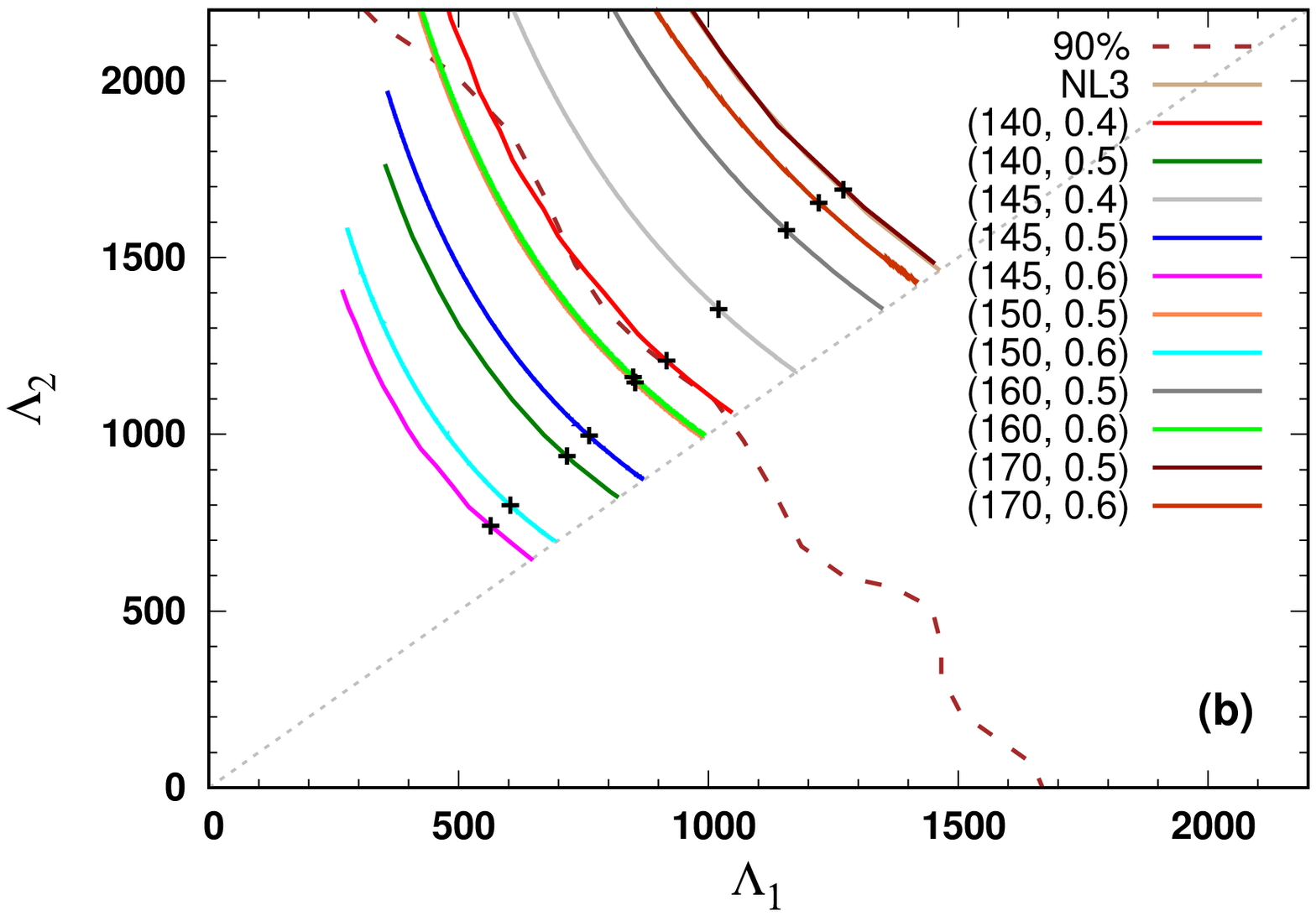}\\
    \includegraphics[width=0.45\textwidth]{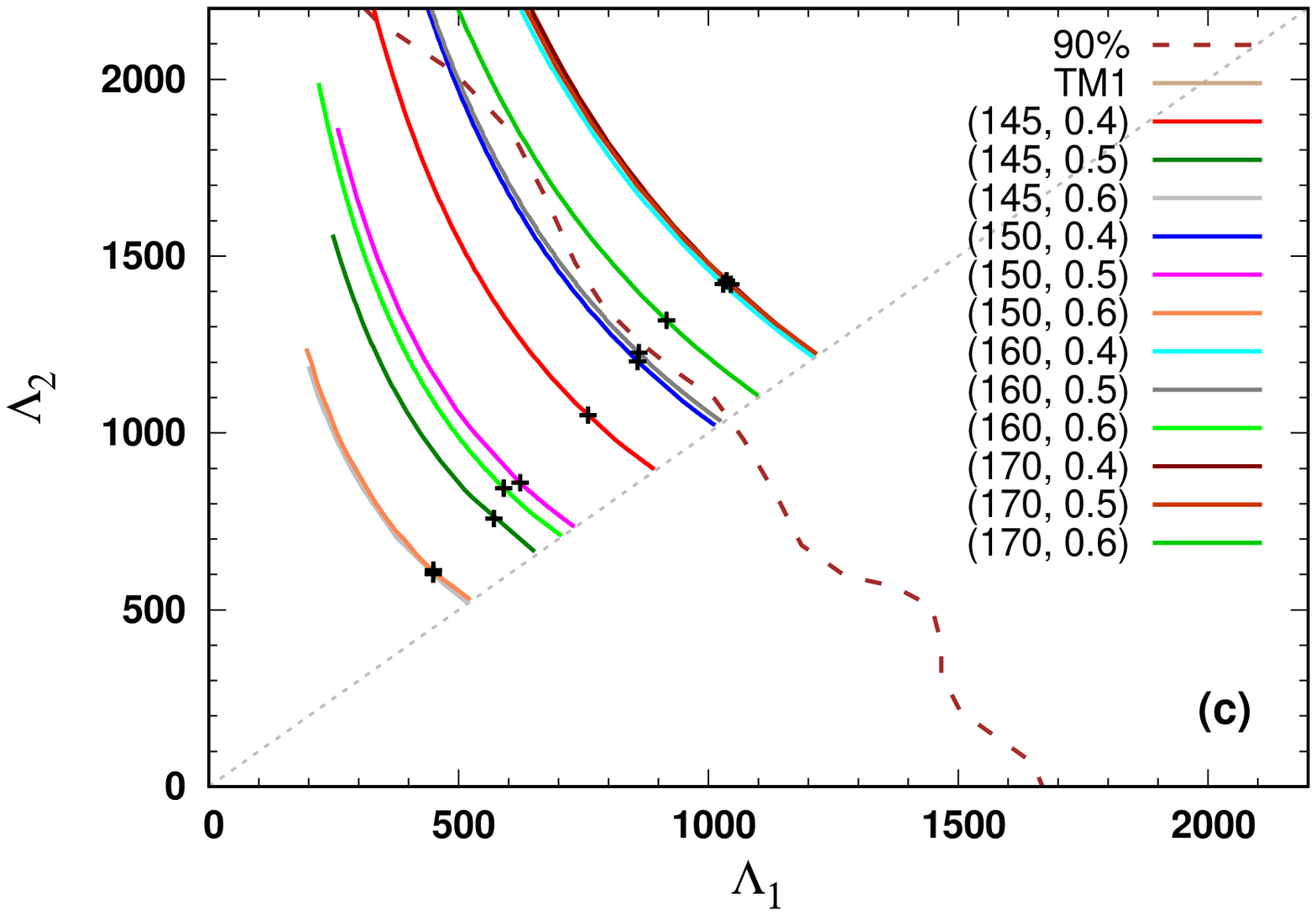}&
    \includegraphics[width=0.45\textwidth]{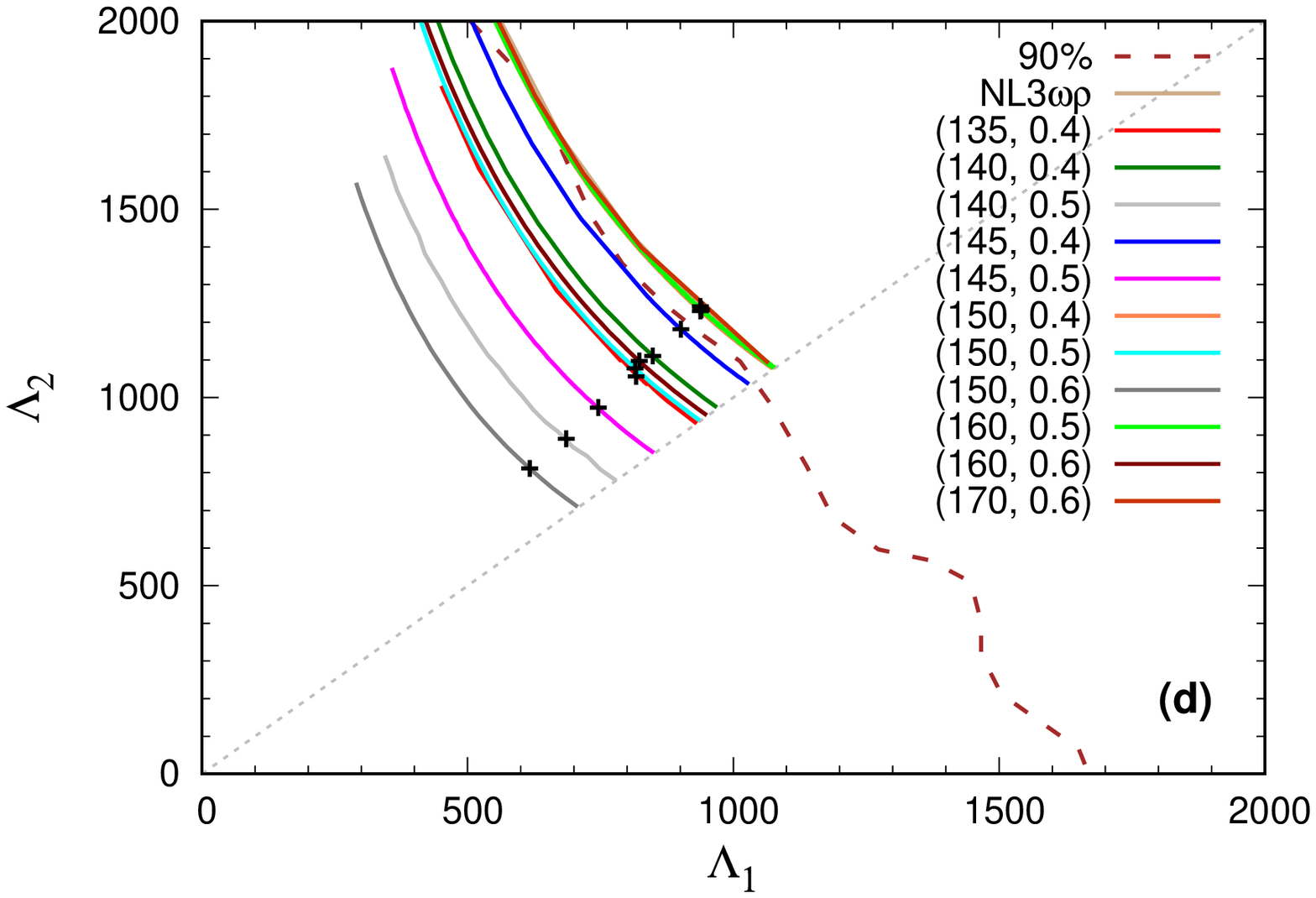}\\
  \end{tabular}
  \caption{Tidal deformabilities $\Lambda_1$ and $\Lambda_2$ corresponding to the high mass $M_1$ and low mass $M_2$ components of the binary system for the event GW170817 calculated for (a) pure hadronic stars and hybrid stars with (b)NL3, (c)TM1 and (d)NL3$\omega\rho$ parameter sets.}
\label{fig:L1-L2}
\end{figure}
Next, we calculate the tidal deformabilities of all the EOS of hybrid stars as well as pure hadronic stars considered here. In Fig. \ref{fig:L1-L2} we plot the individual tidal deformabilities $\Lambda_1$ and $\Lambda_2$ of both the compact stars associated with the binary merger event GW170817. For a given EOS, we calculate $\Lambda_1$ and $\Lambda_2$ by varying the mass of the primary star between $M_1=1.365-1.600M_\odot$ and that of secondary star in the range $M_2=1.170-1.365M_\odot$, so that the combinations of $M_1$ and $M_2$ give the chirp mass ${\cal M}=(M_1M_2)^{3/5}(M_1+M_2)^{-1/5}=1.188M_\odot$ of GW170817 \citep{Abbott2017}. In all the plots of Fig. \ref{fig:L1-L2}, we show the results for only those EOS that gives
$M_{\rm max}$ consistent with the observation \citep{Antoniadis2013}. From the figure it is seen that only NL3$\omega\rho$ EOS is close to the 90\% probability contour set by the event GW170817, by assuming low-spin priors $|\chi|\leq 0.05$ \citep{Abbott2017}. Other two EOS (NL3 and TM1) lie far from the contour and are therefore excluded. Interestingly, when PT from hadronic matter to quark matter via Gibbs' construction is included we find several EOS corresponding to different ($B_{\rm eff}^{1/4},a_4$) combinations satisfy the constraint set by the contour, for all three hadronic models. 

In Fig. \ref{fig:L1-L2} we also mark the points ($\Lambda_1,\,\Lambda_2$) that correspond to $M_1=1.4M_\odot$, with cross signs. Careful observation reveals that not all the EOS which are inside the 90\% contour satisfy the constraint $\Lambda(1.4M_\odot)\leq800$ derived for the event GW170817  by \citet{Abbott2017}.
\begin{table}
\centering
\caption{Maximum allowed values of $B_{\rm eff}^{1/4}$ and $a_4$ obtained from Fig. \ref{fig:Mmax}.}
\label{tab:parameter}
\begin{tabular}{ccc}
\hline 
Model & $B_{\rm eff}^{1/4}$ (MeV) & $a_4$ \\\hline\hline
NL3&160&0.62\\
TM1&162&0.55\\
NL3$\omega\rho$&163&0.65\\
\hline
\end{tabular}
\end{table}
To see the consequence of this constraint on the EOS of hybrid stars we include it (showed by long arrow) in all the plots of Fig. \ref{fig:Mmax}. It is remarkable to find that the constraint reduce the $(B_{\rm eff}^{1/4},a_4)$ parameter space (marked with "allowed" in the plots) significantly in all three cases. The maximum allowed values for $B_{\rm eff}^{1/4}$ and 
$a_4$ are given in Table \ref{tab:parameter}. It is also observed that for NL3 and NL3$\omega\rho$, the core of the maximum mass
configuration mostly contains pure quark matter, whereas for TM1, the core contains mixed phase for the allowed values of
$B_{\rm eff}^{1/4}$ and $a_4$.

\begin{figure}
\centering
\begin{tabular}{ll}
\includegraphics[width=0.36\textwidth,angle=-90]{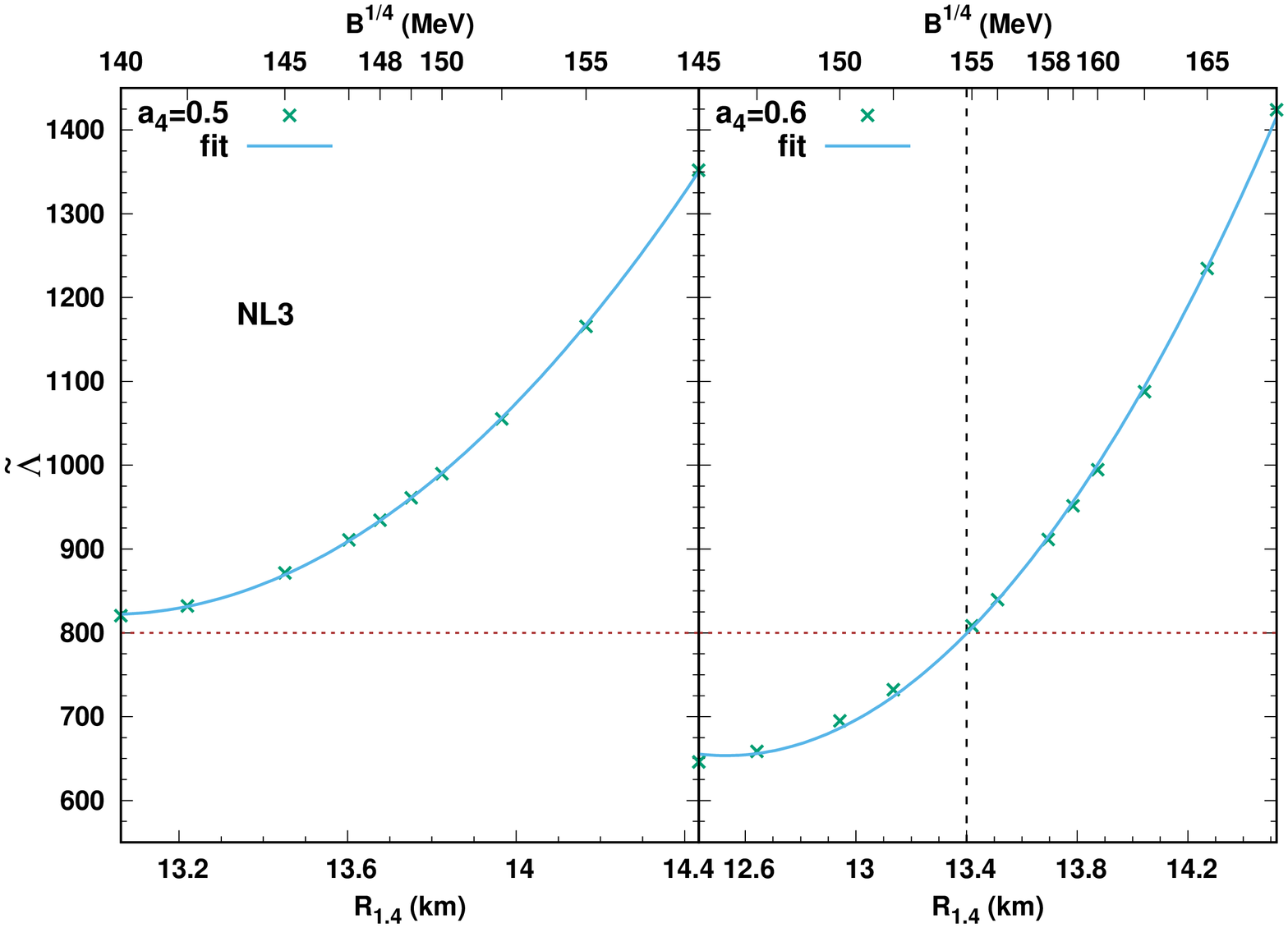}&
\includegraphics[width=0.36\textwidth,angle=-90]{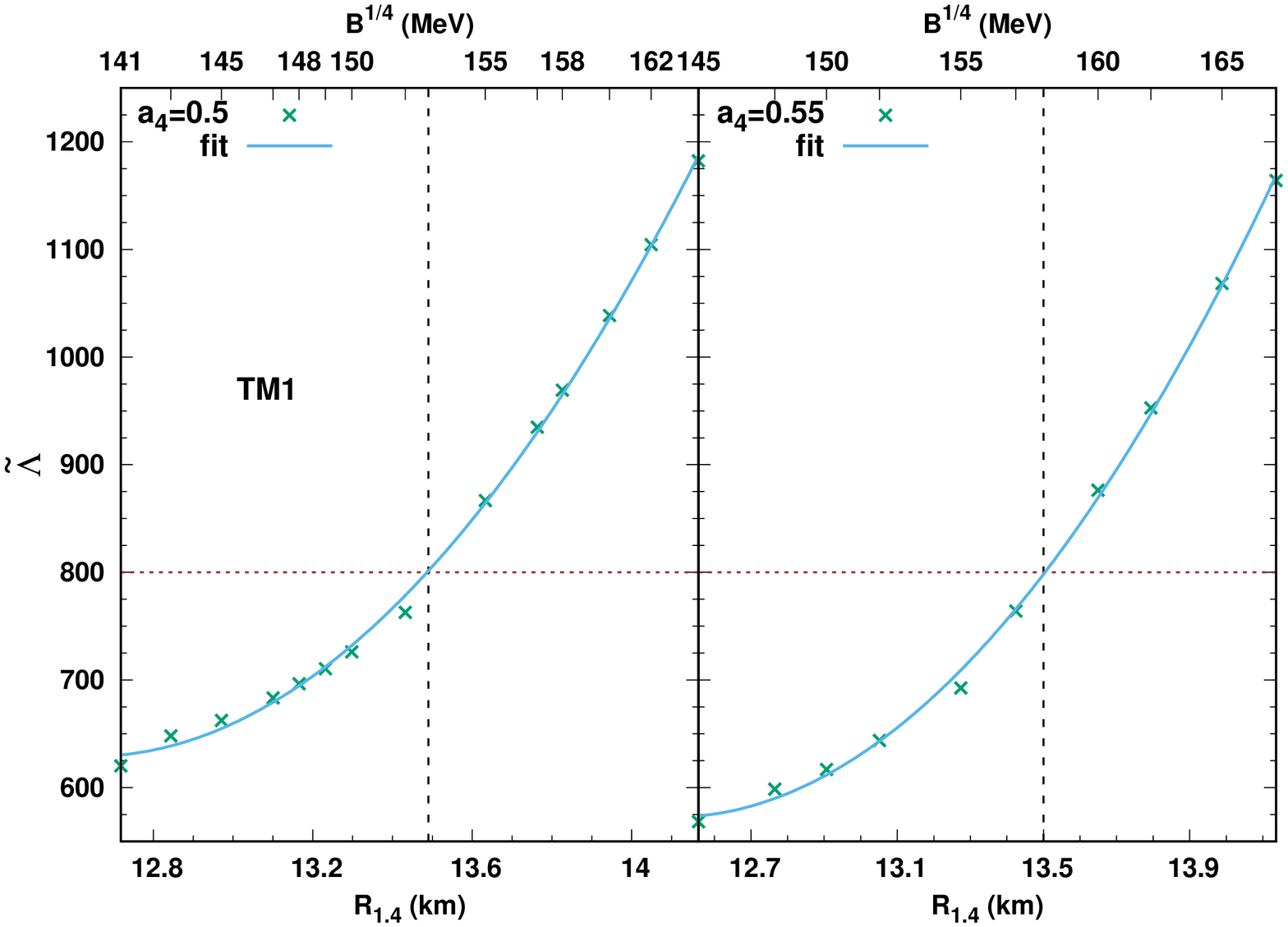}\\
\end{tabular}
\includegraphics[width=0.36\textwidth,angle=-90]{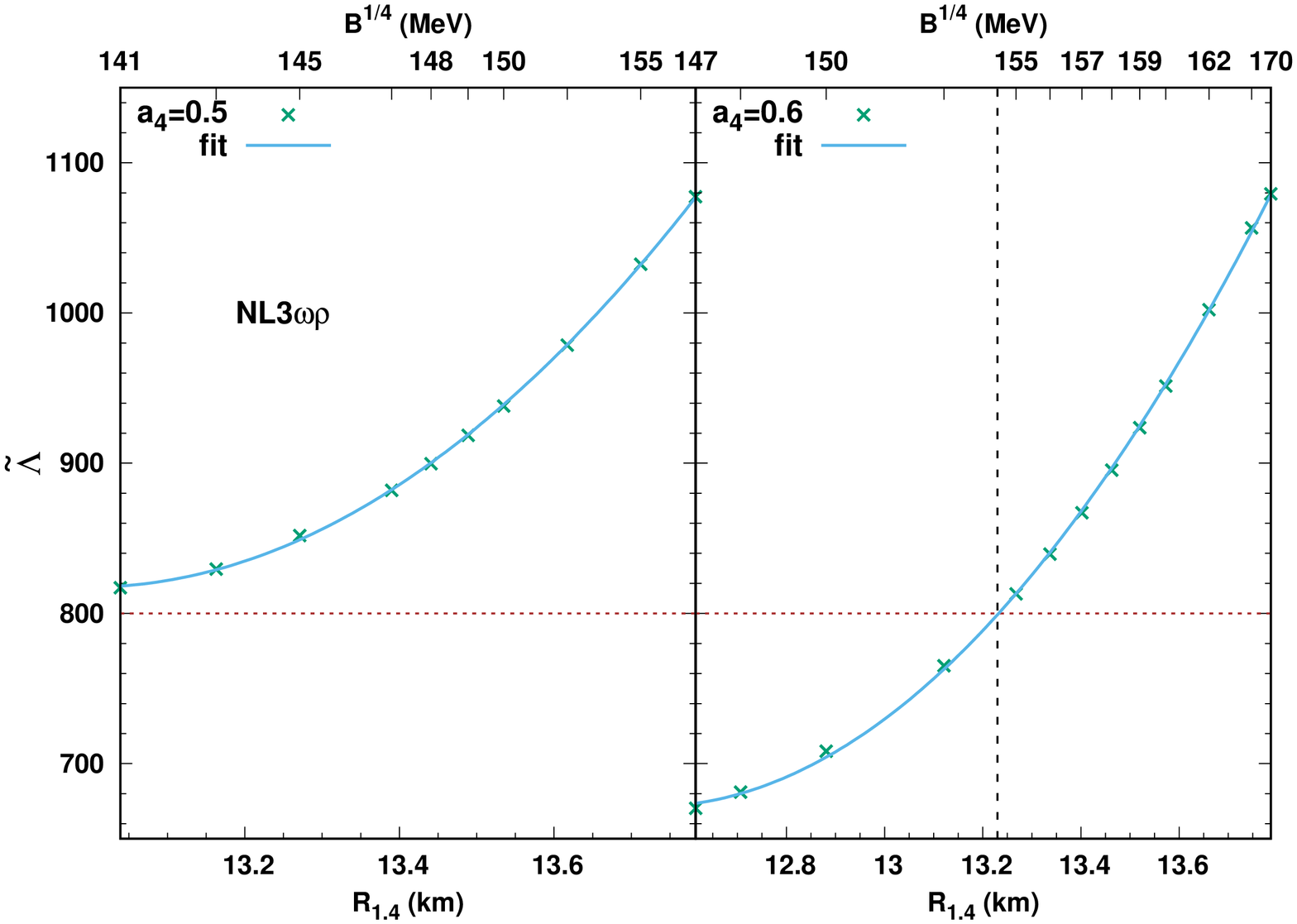}
\caption{The calculated values of $\tilde{\Lambda}$ for the equal mass scenario as a function $R_{1.4}$ (lower x-axis) and
$B_{\rm eff}^{1/4}$ (upper x-axis) for parameter sets NL3, TM1 and NL3$\omega\rho$. 
The solid line is the fit given by Eq. (\ref{eq:fit}). The horizontal dashed line at $\tilde{\Lambda}$=800 is the constraint
from GW170817 \citep{Abbott2017}. The vertical line represents the upper bound on $R_{1.4}$ and $B_{\rm eff}^{1/4}$ due to this 
constraint.}
\label{fig:R14}
\end{figure}
\begin{table}
\centering
\caption{Parameters for fit of $\tilde{\Lambda}$ as a function of $R_{1.4}$ and resulting bounds.}\label{tab:fit_R14}
\begin{tabular}{cccccc}
\hline
\multicolumn{6}{c}{$M_1=M_2=1.365M_\odot$}\\ \hline\hline
Model & $a$ & $b$ & $c$ &  $R_{1.4}$(km) &$B_{\rm eff}^{1/4}$(MeV)\\ \hline 
NL3,\,$a_4=0.50$&46517.4&-7012.76&269.057&\multicolumn{2}{c}{no solution}\\
NL3,\,$a_4=0.60$&30816.9&-4814.86&192.145&13.40&154.7\\
TM1,\,$a_4=0.50$&38963.5&-6064.57&239.855&13.49&153.0\\
TM1,\,$a_4=0.55$&34512.0&-5438.14&217.838&13.50&158.0\\
NL3$\omega\rho$,\,$a_4=0.50$&71678.7&-10904.6&419.515&\multicolumn{2}{c}{no solution}\\
NL3$\omega\rho$,\,$a_4=0.60$&40527.3&-6366.3&254.227&13.23&154.5\\
\hline
\end{tabular}
\end{table}
Next, we investigate the effect of the combined tidal deformability constraint as derived for the event GW170817 assuming low-spin scenario \citep{Abbott2017}, on the EOS of hybrid stars. We consider the case of equal mass with $M_1=M_2=1.365M_\odot$ as well as the case of unequal mass with $M_1=1.60M_\odot,\,M_2=1.17M_\odot$. 
The bound coming from GW170817 is $\tilde{\Lambda}\leq800$.
In Fig. \ref{fig:R14}, we use this bound for equal mass case to predict the radius ($R_{1.4}$) of a $1.4M_\odot$ star as well as $B_{\rm eff}^{1/4}$ for two allowed values of $a_4$ (see Fig. \ref{fig:Mmax}) for all three parameter sets. 
It is found that for both NL3 and NL3$\omega\rho$ parameter sets, $a_4=0.5$ is, in fact, not allowed as it gives
$\tilde{\Lambda}>800$ , for all values of $B_{\rm eff}^{1/4}$. However, for $a_4=0.6$, $\tilde{\Lambda}\leq800$ constraint
give upper bounds on $R_{1.4}$ as $R_{1.4}\leq13.4$ km and $R_{1.4}\leq13.2$ km, respectively.  On the other hand, for TM1
both $a_4=0.5$ and 0.55 are allowed and predict similar upper bound as $R_{1.4}\leq13.5$ km. 
These values of $R_{1.4}$ and corresponding bounds on $B_{\rm eff}^{1/4}$ are listed in the last two columns of 
Table \ref{tab:fit_R14}. The values of $\tilde{\Lambda}$ can easily be fitted with a quadratic function of the form:
\begin{equation}
\tilde{\Lambda}_{\rm fit}=a+bR_{1.4}+cR_{1.4}^2.\label{eq:fit}
\end{equation}
The fit parameters are given in the first three columns of Table \ref{tab:fit_R14}. We include these fits in Fig. \ref{fig:R14}. 
The predicted bounds for $R_{1.4}$ can also be calculated by putting
$\tilde{\Lambda}=800$ in eq. (\ref{eq:fit}) and the obtained values are in good agreement with the values obtained from 
the figures (Fig. \ref{fig:R14}). It is seen from the Table \ref{tab:fit_R14} 
that for TM1 parameter set, the upper bound on $B_{\rm eff}^{1/4}$ increases with $a_4$, unlike the case of $R_{1.4}$ .
For $a_4=0.6$, NL3 and NL3$\omega\rho$ predict similar limiting values for $B_{\rm eff}^{1/4}<155$ MeV. 

\begin{table}
\centering
\caption{Fit parameters and bounds for unequal mass scenario. }
\label{tab:fit_R16}
\begin{tabular}{cccccc}\hline
\multicolumn{6}{c}{$M_1=1.60M_\odot$, $M_2=1.17M_\odot$}\\ \hline\hline
Model & $a$ & $b$ & $c$ &  $R_{1.6}$(km) &$B_{\rm eff}^{1/4}$(MeV)\\ \hline
NL3,\,$a_4=0.50$&53151.5&-8069.29&310.702&13.35&142.0\\ 
NL3,\,$a_4=0.60$&25530.6&-4095.41&168.004&13.35&154.6\\
TM1,\,$a_4=0.50$&55505.7&-8841.12&355.71&13.23&152.1\\
TM1,\,$a_4=0.55$&35994.5&-5914.56&246.207&13.16&157.0\\
NL3$\omega\rho,\,a_4=0.50$&45123.8&-6926.73&270.201&13.32&143.2\\
NL3$\omega\rho,\,a_4=0.60$&17894.3&-2931.8&123.759&13.32&154.6\\
\hline
\end{tabular}
\end{table}
\begin{figure}
\centering
\begin{tabular}{cc}
\includegraphics[width=0.36\textwidth,angle=-90]{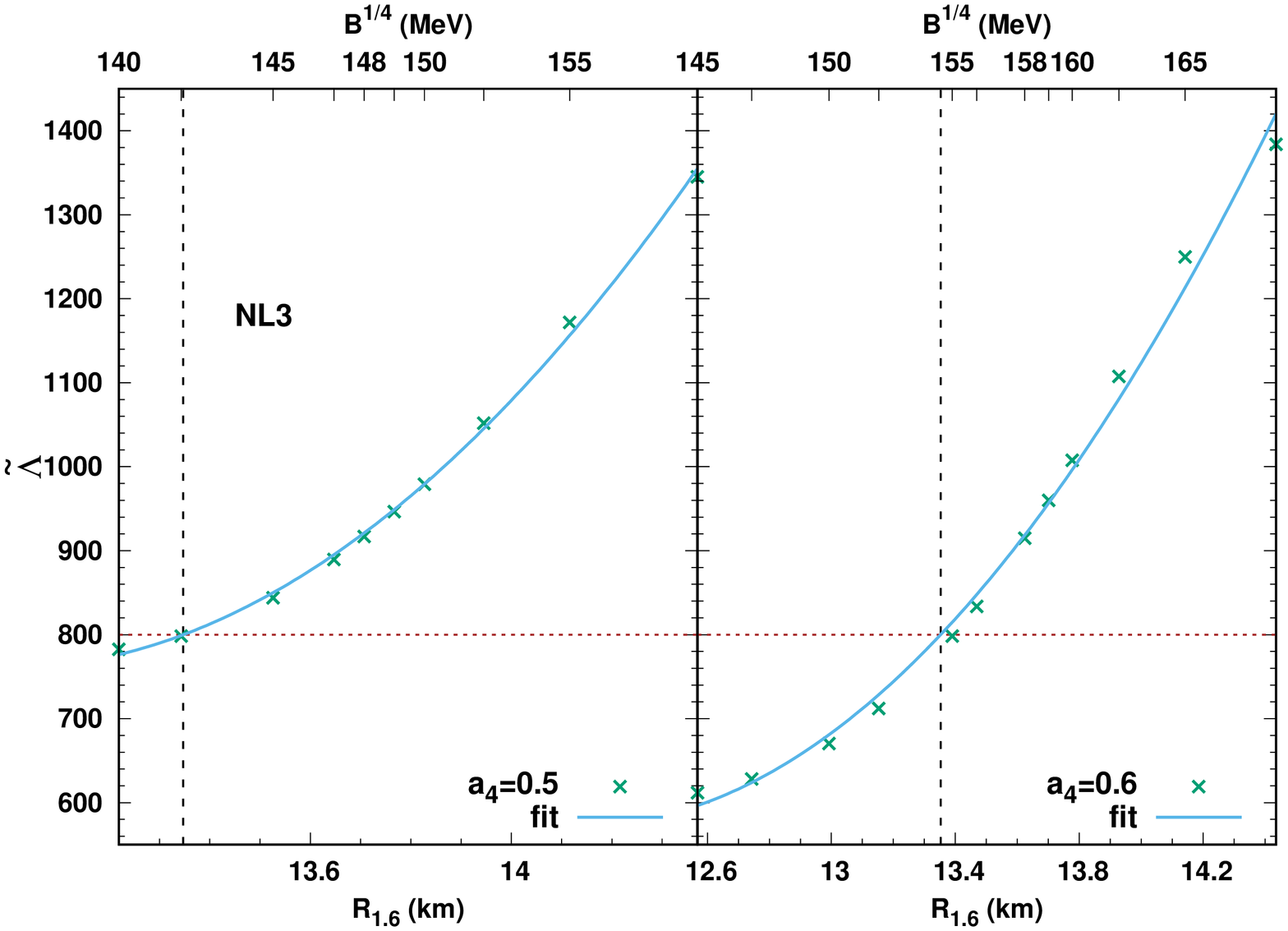}&
\includegraphics[width=0.36\textwidth,angle=-90]{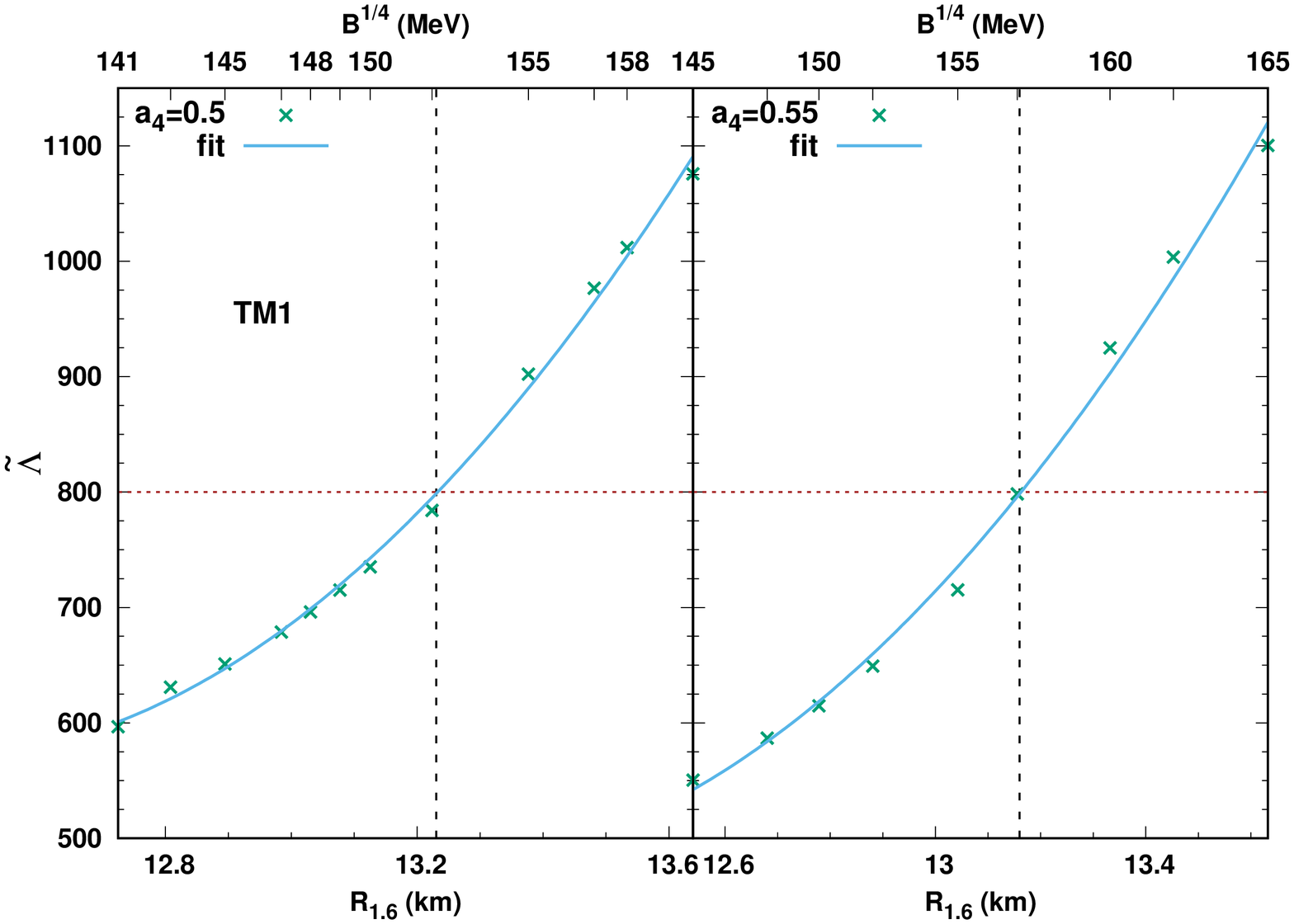}\\
\end{tabular}
\includegraphics[width=0.36\textwidth,angle=-90]{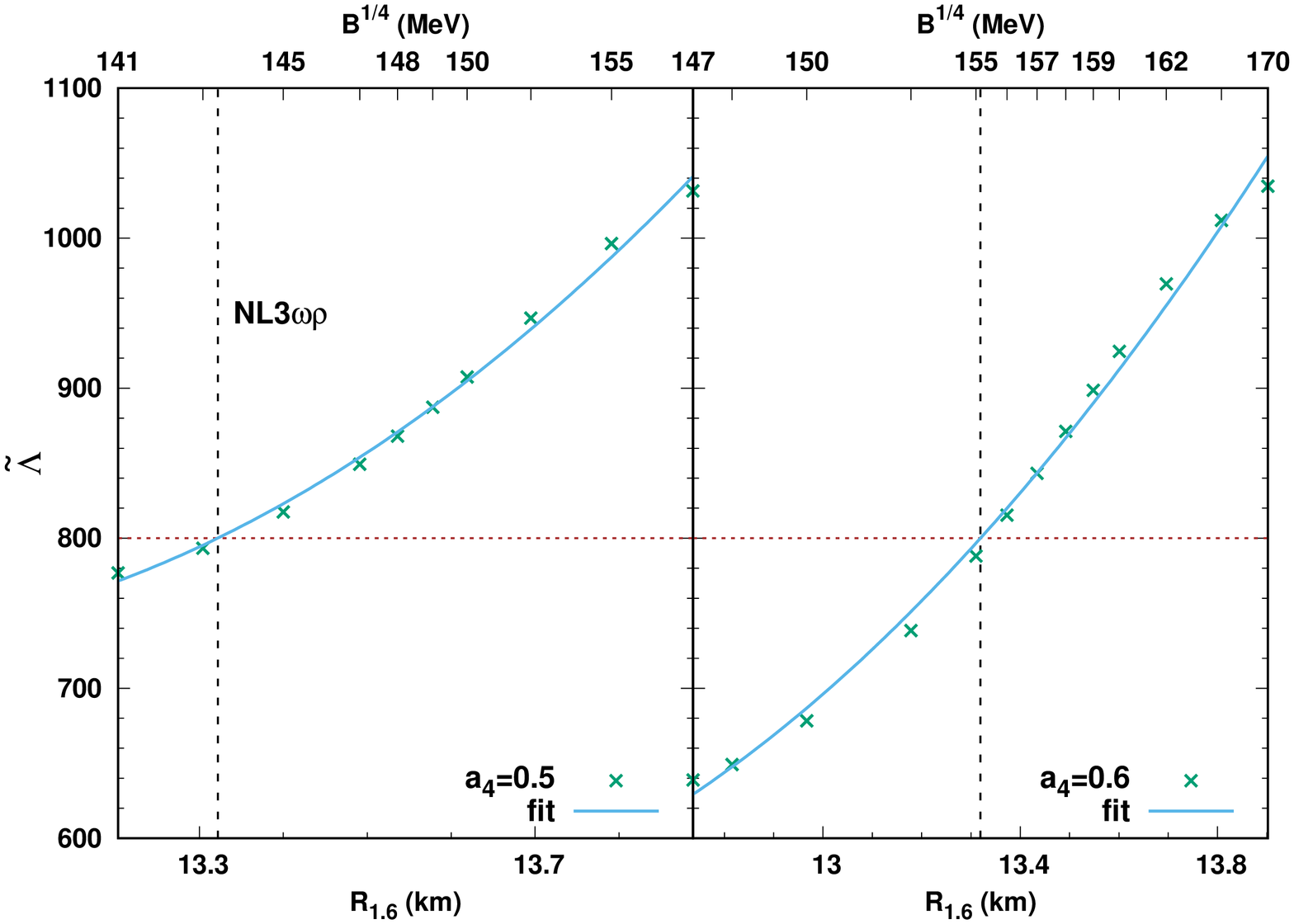}
\caption{Similar to Fig. \ref{fig:R14} but for the unequal mass scenario and as a function of radius of a $1.6M_\odot$ star $R_{1.6}$ (lower x-axis). The horizontal dashed line at $\tilde{\Lambda}$=800 is the constraint from GW170817 \citep{Abbott2017}. The vertical line represents the corresponding upper bound on $R_{1.6}$ and $B_{\rm eff}^{1/4}$.} 
\label{fig:R16}
\end{figure}
We perform similar analysis for the unequal mass case and obtain bound on $R_{1.6}$, the radius of a $1.6M_\odot$ star and $B_{\rm eff}^{1/4}$. The plots including the fits are shown in Fig. \ref{fig:R16}. The fit parameters and the resulting bounds from the $\tilde{\Lambda}=800$ constraint are given in Table \ref{tab:fit_R16}.
In this case both the values of $a_4$ are found to be allowed for all three parameter sets. The bounds on $R_{1.6}$ are almost independent of the value of $a_4$ and all three parameter sets predict similar bounds for $R_{1.6}$ as $R_{1.6}\lesssim 13.2-13.4$ km. The corresponding limits on $B_{\rm eff}^{1/4}$ are found to be similar for NL3 and NL3$\omega\rho$
and a few MeV larger for TM1.

When we combine the lower bound on $B_{\rm eff}^{1/4}$ set by the  requirement that $n_{\rm crit}>n_{\rm cctd}$ with the upper bounds on $B_{\rm eff}^{1/4}$ obtained above (Figs. \ref{fig:Mmax}, \ref{fig:R14} and \ref{fig:R16}) using the constraints on tidal deformability coming from the event GW170817, we get the range of values for $B_{\rm eff}^{1/4}$ given in the Table \ref{tab:beff}.
We have also included the corresponding starting densities of mixed phase ($n_{\rm crit}$) in the table.
It is clear from the table that both the equal and unequal mass scenarios predict similar upper bounds on $B_{\rm eff}$. 
This is in agreement with the earlier finding by \citet{Radice2018}, that the combined tidal deformability $\tilde{\Lambda}$ is
almost independent of $q=M_2/M_1$. The $\tilde{\Lambda}\leq 800$ constraint is found to set more stringent limit 
on $B_{\rm eff}$ than the $\Lambda(1.4M_\odot)\leq800$ constraint. It is also observed that for all three parameter sets the
hadron-quark mixed phase should start relatively early ($n_{\rm crit}\sim 0.1-0.2$ fm$^{-3}$). 
\begin{table}
\caption{Range of allowed values of $B_{\rm eff}^{1/4}$ and corresponding $n_{\rm crit}$.}
\label{tab:beff}
\begin{tabular}{c|c|c|c|c}
\hline 
$a_4$&\multicolumn{4}{c}{$B_{\rm eff}^{1/4}$ (MeV), $n_{\rm crit}$ (fm$^{-3}$) }\\
\hline
&lower bound & \multicolumn{3}{c}{upper bound}\\
\hline 
constraint& $n_{\rm crit}>n_{\rm cctd}$&$\tilde{\Lambda}\leq800$& $\tilde{\Lambda}\leq800$&$\Lambda(1.4M_\odot)\leq800$\\

&&({\small $M_1=M_2=1.365M_\odot$)}&({\small $M_1=1.60M_\odot,\,M_2=1.17M_\odot$})&\\
\hline\hline 
\multicolumn{5}{c}{NL3 ($n_{\rm cctd}=0.056$ fm$^{-3}$)}\\
\hline 
0.50 & $138$&$-$&$142,\,0.102$&$148,\,0.142$\\
0.60 & $145$&$155,\,0.146$&$155,\,0.146$&$159,\,0.174$\\
\hline 
\multicolumn{5}{c}{TM1 ($n_{\rm cctd}=0.058$ fm$^{-3}$)}\\
\hline\hline 
0.50& $141$&$153,\,0.178$&$152,\,0.170$&$158,\,0.215$\\
0.55& $142$&$158,\,0.191$&$157,\,0.183$&$162,\,0.220$\\
\hline 
\multicolumn{5}{c}{NL3$\omega\rho$ ($n_{\rm cctd}=0.082$ fm$^{-3}$)}\\
\hline\hline
0.50 & $140$&$-$&$143,\,0.116$& $149,\,0.179$\\
0.60 & $147$&$155,\,0.176$&$155,\,0.176$& $159,\,0.214$\\
\hline 
\end{tabular}
\end{table}

The first model-independent measurement of neutron skin thickness is provided by the Lead Radius Experiment PREX as $R_{\rm skin}^{208}=0.33_{-0.18}^{+0.16}$ fm \citep{Abrahamyan2012}. The result is not very useful because of the large statistical error. However, the upcoming PREX-II experiment is supposed to reduce the uncertainty to $0.06$ fm. On the other hand, the calculated values of $R_{\rm skin}^{208}$ within NL3, TM1 and NL3$\omega\rho$ parameter sets are 0.28 fm, 0.27 fm and 0.21 fm, respectively. If PREX-II gives a value closer to the current central value of 0.33 fm NL3$\omega\rho$ would be excluded. However, NL3 and TM1 might be consistent with the measured $R_{\rm skin}^{208}$ as well as the bound set by the event GW170817 only if the PT to quark matter is taken into consideration, as already indicated by \citet{Fattoyev2017}.

\section{Summary}\label{sec:summary}
We have investigated here how the tidal deformability bound given by the binary NS merger event GW170817 constraints the EOS of
hybrid stars. The EOS of hybrid stars are constructed via the Gibbs' construction through the formation of hadron-quark mixed
phase. For the hadronic part we employ the RMF model with three widely used parameter sets: NL3, TM1 and NL3$\omega\rho$, whereas
the quark matter is described with the help of the modified bag model. It is found that pure hadronic EOS for all three 
parameters sets are excluded as they give values of $\Lambda$ considerably larger than the constraint
$\Lambda(1.4M_\odot)\leq800$, set by GW170817. They also lie outside the 90\% contour in the ($\Lambda_1, \Lambda_2$) plane. 
Interestingly, when the PT to quark matter via the formation of mixed phase is considered, we find several combinations of
$B_{\rm eff}^{1/4}$ and $a_4$ that are consistent with the above constraints for all three cases. However, imposition of
constraints $\Lambda(1.4M_\odot)\leq800$ and $M_{\rm max}=2.01\pm0.04M_\odot$ together leads to significant reduction of
($B_{\rm eff}^{1/4},\,a_4$)  parameter space.
We have also investigated the effect of combined tidal deformability constraints $\tilde{\Lambda}\leq800$ derived from the event 
GW170817 assuming low spin priors. This bound is found to put stringent constraints on the allowed values of $B_{\rm eff}^{1/4}$
and $a_4$. We have obtained  bound for radius of a $1.4M_\odot$ hybrid stars as $R_{1.4}\leq 13.4$ km for NL3, 
$R_{1.4}\leq 13.5$ km for TM1 and $R_{1.4}\leq 13.2$ km for NL3$\omega\rho$. Similarly, bounds on the radius of 
a $1.6M_\odot$ star are obtained as $R_{1.6}\leq13.4$ km for NL3, $R_{1.6}\leq13.2$ km for TM1 and $R_{1.6}\leq 13.3$ km 
for NL3$\omega\rho$. We also observed that if the upcoming PREX-II experiment measures relatively larger Pb$_{\rm skin}^{208}$,
EOS of hybrid stars with NL3$\omega\rho$ would be excluded but NL3 and TM1 hybrid stars would survive.  
\begin{acknowledgements}
P. Char acknowledges support from the Navajbai Ratan Tata Trust.
\end{acknowledgements}

\end{document}